\documentclass[journal]{IEEEtran}
\usepackage{amsmath,amsfonts}
\usepackage{algorithmic}
\usepackage{algorithm}
\usepackage{array}
\usepackage[caption=false,font=normalsize,labelfont=sf,textfont=sf]{subfig}
\usepackage[table,xcdraw]{xcolor}
\usepackage{textcomp}
\usepackage{stfloats}
\usepackage{url}
\usepackage{verbatim}
\usepackage{graphicx}
\usepackage{cite}
\begin{document}

\title{An Embedded Auto-Calibrated Offset Current Compensation Technique for PPG/fNIRS System}

\author{{Sadan Saquib Khan, Sumit Kumar, Benish Jan, Laxmeesha Somappa, and Shahid Malik}
\thanks{Sadan Saquib Khan, Sumit Kumar, Benish Jan, and Shahid Malik are with the Centre for Sensors, Instrumentation and Cyber Physical System Engineering (SeNSE), Indian Institute of Technology Delhi (IIT Delhi).}
\thanks{Laxmeesha Somappa is with the Department of Electrical Engineering, Indian Institute of Technology Bombay (IIT Bombay). }
\thanks{Manuscript received Month Day, Year; revised Month Day, Year.}}

\markboth{IEEE Embedded Systems Letters,~Vol.~XX, No.~YY, Month~Year}%
{Shell \MakeLowercase{\textit{et al.}}: A Sample Article Using IEEEtran.cls for IEEE Journals}


\maketitle

\begin{abstract}
Usually, the current generated by the photodiode proportional to the oxygenated blood in the photoplethysmography (PPG) and functional infrared spectroscopy (fNIRS) based recording systems is small as compared to the offset-current. The offset current is the combination of the dark current of the photodiode, the current due to ambient light, and the current due to the reflected light from fat and skull \cite{b1,b2}. The relatively large value of the offset current limits the amplification of the signal current and affects the overall performance of the PPG/fNIRS recording systems. In this paper, we present a mixed-signal auto-calibrated offset current compensation technique for PPG and fNIRS recording systems. The system auto-calibrates the offset current, compensates using a dual discrete loop technique, and amplifies the signal current. Thanks to the amplification, the system provides better sensitivity. A prototype of the system is built and tested for PPG signal recording. The prototype is developed for a 3.3 V single supply. The results show that the proposed system is able to effectively compensate for the offset current.
\end{abstract}


\section{Introduction}
\IEEEPARstart{T}{he} Optical sensors are used in many applications thanks to their unique features such as immunity to electromagnetic interference, small size, lightweight and high sensitivity \cite{b3, b4, b5}. They are particularly useful in biomedical applications since they non-invasively provide information about the biomarkers. Various biomedical diagnostic devices such as oximeters, cerebral oximeters, and fNIRS-based brain-imaging systems utilize optical sensors \cite{b6, b7, b8}. In addition, they are also being explored for the non-invasive detection of breast cancer \cite{b9,b10,b11}.  However, the optical sensors suffer from the offset current. Mostly a photodiode is used to convert the photon into the current. The offset current may be because of the ambient light, dark current, and also because of the light reflected from the fat, bones, and skulls \cite{b12}. Usually, the dark current is fixed and can be compensated. However, the offset current due to ambient light and reflection varies with time and can be significantly higher than the current due to the actual signal \cite{b13,b14,b15,b16}.

\begin{figure}[!ht]
    \centering
    \includegraphics[trim=15 15 10 10, scale=0.45 ]{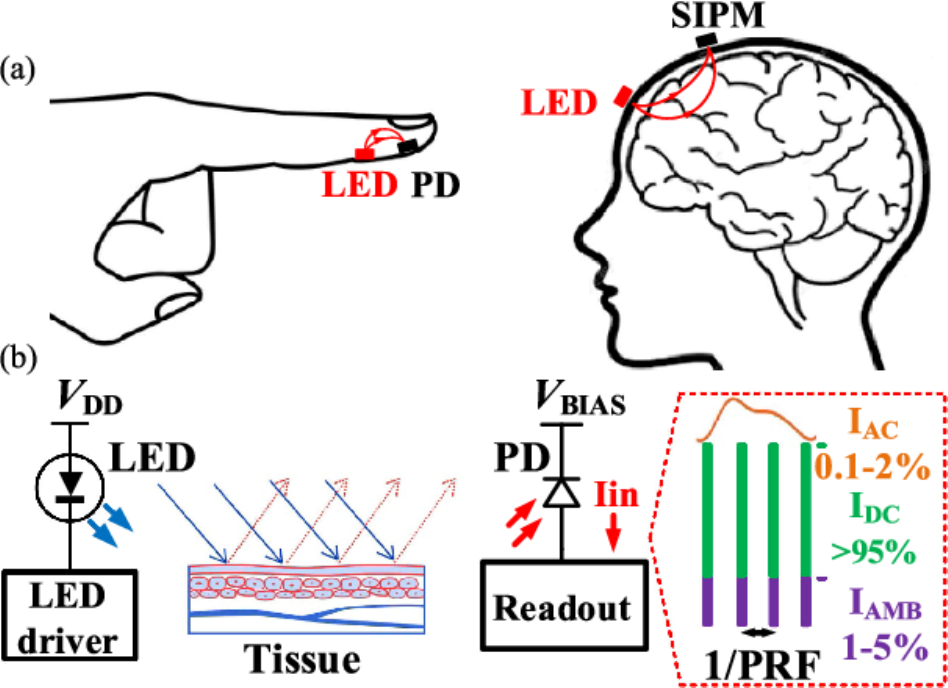}
    \caption{Overview of PPG sensor \cite{b8}.}
    \label{fig:1}
\end{figure}

The current from the photodiode is converted into equivalent voltage using a trans-impedance amplifier. The value of the feedback resistor is selected appropriately to convert the current into an output voltage. A high-value of feedback resistor is desirable to amplify the signal and keep the current noise at a low level. However, the large value of offset current limits the amplification. To compensate for the effect of offset current on the output voltage of the trans-impedance amplifier, various approaches are explored.
The current source using the digital-to-analog converter (DAC) for offset current compensation is integrated with many system-on-chip (SoC) based PPG/fNRIS systems. The current-DAC-based SoC systems are able to compensate wide-range offset current from 1$\mu A$ to 128 $\mu A$ with a 7-bit resolution. However, the \\
\textcolor{red}{HPF for offset compensation}

 \begin{figure}[ht]
    \centering
    \includegraphics[trim=40 0 0 10, clip, scale = 0.35]{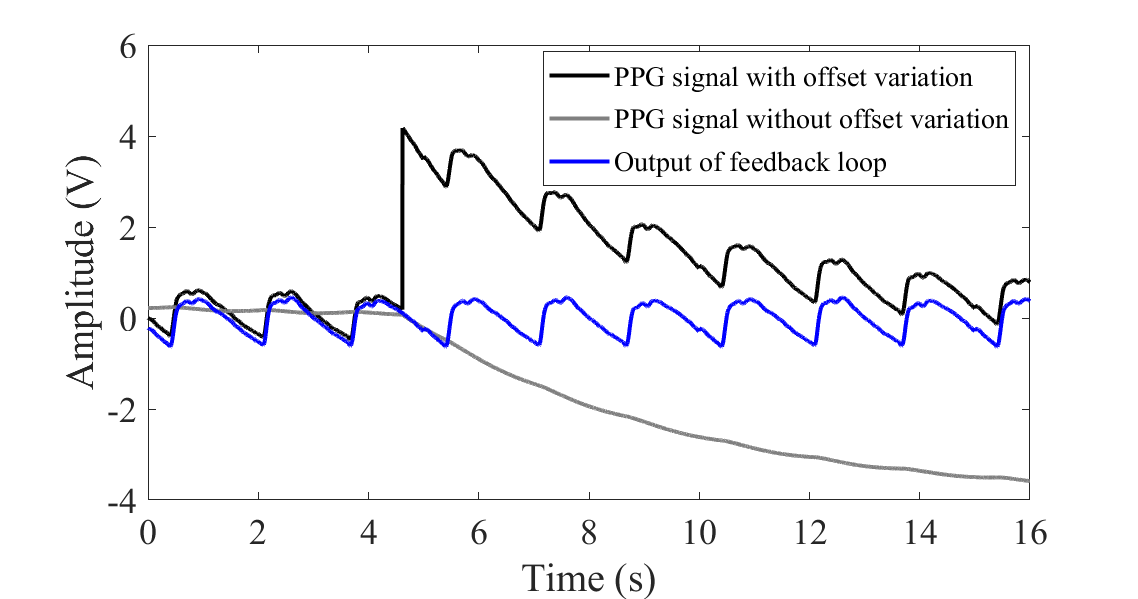}
     \caption{Block diagram of a four-channel impedance measurement unit for impedance sensors} 
   \label{fig:flow}
\end{figure}
 \begin{figure*}[ht]
    \centering
    \includegraphics[trim=20 10 0 20, clip, scale = 0.96]{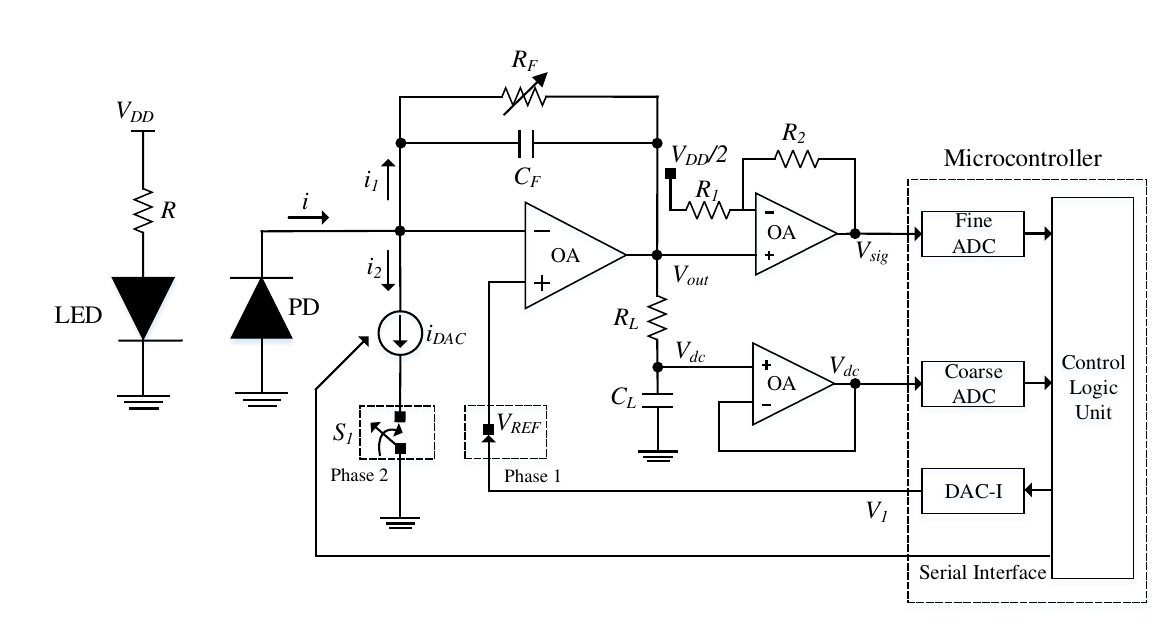}
     \caption{Block diagram of a four-channel impedance measurement unit for impedance sensors} 
   \label{fig:bd}
\end{figure*}

The offset current compensation techniques, based on a negative feedback loop (analog or digital), by nullifying the DC-voltage at the output of the amplifier, are reported in
\cite{b17,b18,b19,b20}. This continuous nullification of the DC signal affects the shape of the signal which is misinterpreted as the measurand. The problem is especially critical for biomedical signals such as a PPG and functional near-infrared spectroscopy (fNIRS) for brain-computer interfacing applications. In addition, the implementation of the analog low-pass filter in the feedback path to nullify the offset current introduces the delay in the sensor signal\cite{b21, b22}. Since the cut-off frequency of the filter is near dc, it introduces a delay causing a phase error in the PPF and fNIRS signals \cite{b23,b24,b25,b26,b27}. Typically, the cutoff frequency varies from 0.1 Hz to 5 Hz for PPG and fNIRS. Due to the extremely low cutoff frequency, the analog low pass filter may influence the frequencies present in our signal of interest \cite{b28,b29,b30,b31}.

In this paper, we present a dual-loop auto-compensation technique for wide-range offset current compensation. The circuit utilizes voltage feedback to compensate for the offset current smaller than 1$\mu A$. For a wide range of currents, the circuit utilizes a digitally controlled current source \cite{b32,b33,b34,b35}. The circuit compensates the offset current.

\section{The Proposed System}
The block diagram of the proposed discrete offset compensation loop-based system is shown in Fig. \ref{fig:bd}. The system consists of a trans-impedance amplifier, with a digital potentiometer $R_F$ for programmable gain. A programmable current source $i_{DAC}$  is used to compensate for the offset current from the photodiode. The proposed system utilizes a dual-cancellation technique to effectively compensate the offset voltage from the output voltage $V_{out}$ of the trans-impedance amplifier. A low-pass filter is used to filter out the DC offset from the voltage $V_{out}$. The DC-offset voltage $V_{dc}$ is acquired using an analog-to-digital converter. The offset-free output is further amplified by a variable gain non-inverting amplifier $OA_2$. The output voltage $V_sig$ of the amplifier is proportional to the AC signal.

The frequency of the PPG and fNIRS is usually between 0.05 Hz to 5 Hz. The selection of an appropriate cut-off frequency for the compensation of DC-offset is important for the effective design of the overall embedded sensing system. For instance, ideally, a cut-off frequency of less than 0.05 Hz is preferred for DC-offset compensation. However, the compensation time will increase accordingly. The continuous compensation technique will also affect the shape morphology of the PPG/fNIRS signals. In this paper, the cut-off frequency of the low-pass filter is chosen to be around 1 Hz. The proposed system utilizes a discrete offset compensation approach by monitoring the offset voltage. The loop will automatically compensate the offset voltage if the DC voltage crosses a threshold voltage limit.

 \begin{figure}[ht]
    \centering
    \includegraphics[trim=0 15 0 18, clip, scale = 0.96]{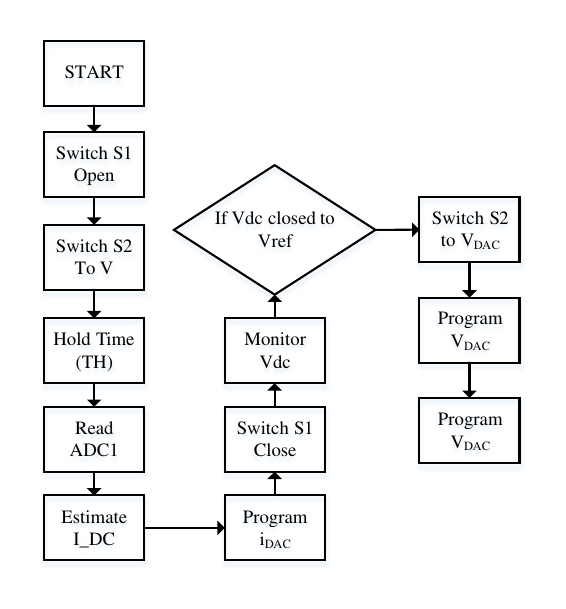}
     \caption{Block diagram of a four-channel impedance measurement unit for impedance sensors} 
   \label{fig:flow1}
\end{figure}

The flow of operation of the proposed system is shown in Fig. \ref{fig:flow}. The auto-calibration phase start with switching the opening of the switch $S_1$ and connecting the switch $S_2$ to $V_{cm}$. During this phase, both ac and dc current through the photo-diode is amplified by the trans-impedance amplifier. The average dc value of $V_{out}$ is extracted using the low-pass filter. The voltage $V_{dc}$ is acquired using an analog-to-digital converter (ADC 1), as shown in Fig. \ref{fig:bd}. The DC current is estimated from the voltage $V_{dc}$ and resistor $R_F$. Next, the magnitude of the digitally controlled current source ($i_{DAC}$) is tuned to compensate for the DC current. Once the current value is set, the switch $S_1$ is closed. Consequently, the DC offset current from the photo-diode has sunk by the $i_{DAC}$. The output voltage of the trans-impedance amplifier is proportional to the AC signal. 

The DC offset is compensated by using the digitally controlled current source. The magnitude of the current source is controlled by using a digital potentiometer. However, the digital potentiometer is suffered from limited resolution and tolerance. Once the value of $i_{DAC}$ is set to compensate for the DC offset, the gain of the trans-impedance amplifier is increased by using digital-potentiometer $R_F$. However, due to the finite resolution of the current source, the uncompensated DC current is also amplified, which results in a DC offset at the output of the trans-impedance amplifier. To compensate for that, in this letter, we incorporated a second offset-compensation loop by controlling the voltage $V_{REF}$ of the trans-impedance amplifier.   

\begin{figure}[!ht]
    \centering
    \includegraphics[trim=15 15 10 10,clip, scale=0.9 ]{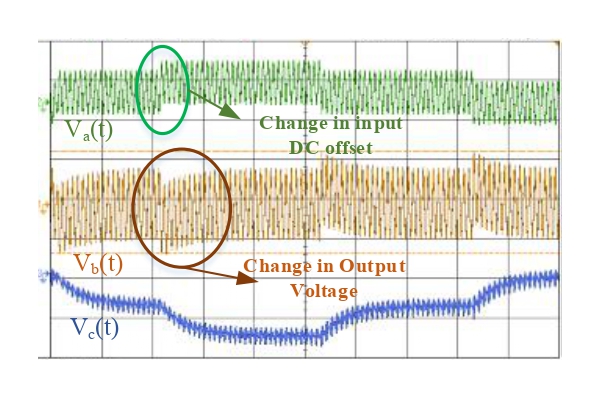}
    \caption{Experimental PPG data acquired with the traditional continuous offset cancellation technique.}
    \label{fig:Figure 4}
\end{figure}
\begin{table}[H]
\centering
\caption{Components used for Prototyping}
\begin{tabular}{|c|c|}
\hline
\rowcolor[HTML]{DAE8FC} 
\textbf{Component}              & \textbf{Details}                                                                  \\ \hline
\rowcolor[HTML]{FFFFFF} 
LED and Photo-Diode wavelengths & 940nm                                                                             \\ \hline
\rowcolor[HTML]{FFFFFF} 
Operational Amplifier           & OPA333                                                                            \\ \hline
\rowcolor[HTML]{FFFFFF} 
$i_{DAC}$                       & LM334                                                                             \\ \hline
\rowcolor[HTML]{FFFFFF} 
Digital-potentiometer           & AD8252                                                                            \\ \hline
\rowcolor[HTML]{FFFFFF} 
Microcontroller                 & \begin{tabular}[c]{@{}c@{}}ATSAMD21G18\\ (12-bit ADC)\\ (10-bit DAC)\end{tabular} \\ \hline
\rowcolor[HTML]{FFFFFF} 
$R_F$                           & 3.9 k$\Omega$ - 1 M$\Omega$                                                       \\ \hline
\rowcolor[HTML]{FFFFFF} 
$R_L$                           & 20 k$\Omega$                                                                      \\ \hline
\rowcolor[HTML]{FFFFFF} 
$C_L$                           & 10 $\mu$F                                                                         \\ \hline
\end{tabular}
\label{table:components}
\end{table}

\section{Experimental Setup and Results}
A prototype of the proposed embedded auto-calibrated system is fabricated and tested. The components used for the prototyping are tabulated in Table \ref{table:components}. The low-voltage low-power operational amplifier (OPA333) is used to implement the trans-impedance amplifier, voltage-follower, and non-inverting amplifier. A microcontroller with integrated ADC and DAC is used. The resolution of the ADC can be tuned from 12 to 16 bits with oversampling and decimation. A digital potentiometer with 8-bit resolution and 1 M$\Omega$ full-scale value is used. The amplitude of the current source can be tuned from 1 $\mu$A to 10 $m$A. 
\begin{figure}[!ht]
    \centering
    \includegraphics[trim=25 20 10 10,clip, scale=0.66 ]{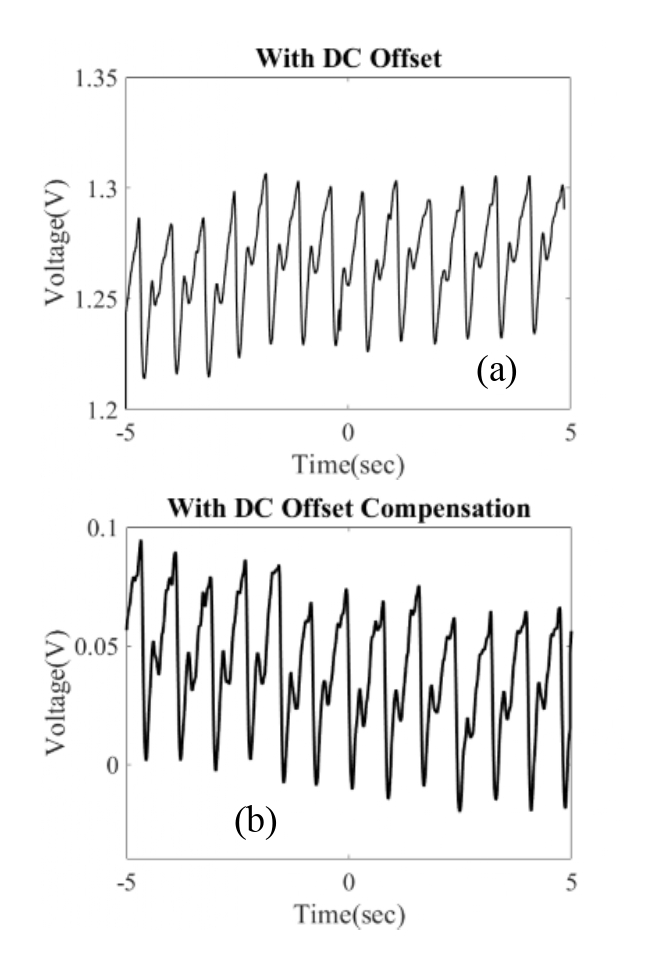}
    \caption{Experimental PPG results obtained with (a) dc offset and with (b) Compensated dc offset.}
    \label{fig:Figure 5}
\end{figure}
This study's primary goal was to look into the DC Offset in the PPG signal. In order to get the PPG signal, infrared light is shone on the fingertip and reflected to a photodiode that has data on the blood and oxygen flow.
From Fig. \ref{fig:Figure 5}a, we can clearly visualize the DC offset present in the signal which limits the gain in the succeeding stage. Fig. \ref{fig:Figure 5} (b) show the dc offset compensated PPG signals by acquiring a stable DC value using the proposed mixed-signal system. The dc offset in the PPG signal has been successfully eliminated up to a very high order with no effect of the delay and shape morphing of the PPG signal.

\section{Conclusion}
A mixed-signal loop based approach for acquiring high fidelity PPG/fNIRS signals has been demonstrated. The proposed architecture overcomes the issues associated with large delay and shape morphing of the signals with the traditional continuous offset cancellation technique. Finally, PPG measurement results acquired with the proposed system are shown and the offset cancellation with high fidelity data acquisition is verified.




\end{document}